\documentclass{appolb}
\usepackage{amsmath}
\usepackage{graphicx}

\def\be{\begin{equation}}
\def\ee{\end{equation}}

\begin{document}
\title{Rapidity gaps and ancestry
  \thanks{Work presented at Diffraction and Low-x 2018, Reggio Calabria, August 2018,
    and supported in part by the Agence Nationale de la Recherche under the project \# ANR-16-CE31-0019.}%
}%
\author{Dung LE ANH, St\'ephane MUNIER
  \address{Centre de physique th\'eorique (CPHT),
    \'Ecole polytechnique, CNRS,\\ Universit\'e Paris-Saclay, Route de Saclay, 91128 Palaiseau, France}
}
\maketitle
\begin{abstract}
  The recently discovered correspondence between the distribution of rapidity gaps in electron-nucleus diffractive processes and
  the statistics of the height of genealogical trees in branching random walks
  is reviewed. In addition, a new comparison of numerical solutions of exact equations for diffraction
  on the one hand, and for ancestry on the other hand, both established
  in the framework of the color dipole model, is presented.
\end{abstract}

\section{Rapidity gap distribution in deep-inelastic scattering}
\label{sec:intro}

\begin{figure}[h]
  \centering
  \includegraphics[width=.7\textwidth,clip]{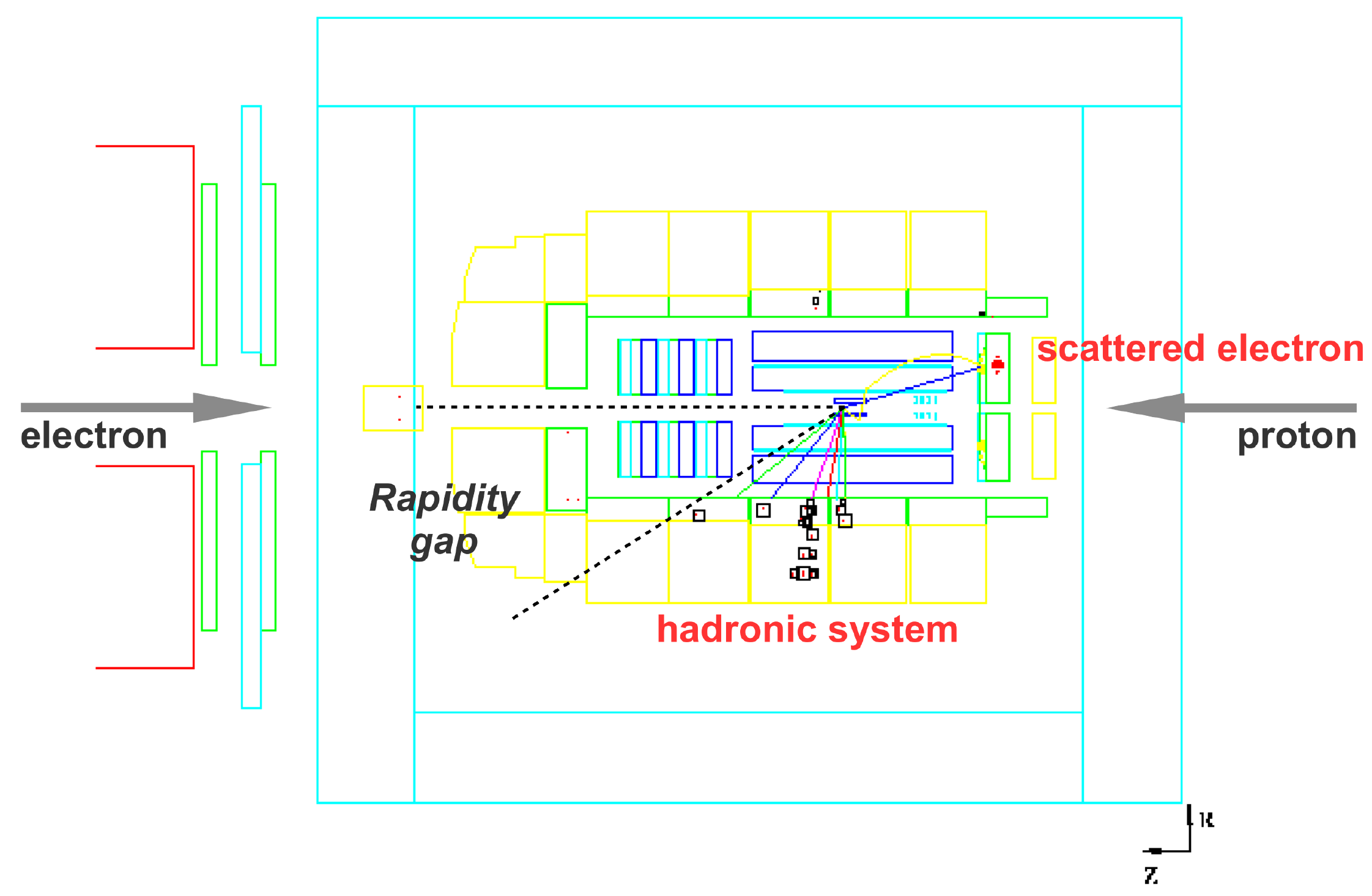}
  \caption{{\it Diffractive dissociation event recorded in the H1 detector.} The highly-energetic proton (entering from the right) interacts elastically, picks a small transverse momentum compared to its longitudinal momentum and therefore does not leave any track in the detector. The virtual photon instead is converted into a hadronic system. The angular sector between the momentum of the scattered proton and the produced hadronic system void of any activity is the {rapidity gap}.}
\label{fig-1}
\end{figure}

In the scattering of electrons off protons at very high energies,\footnote{For background on all aspects of high-energy scattering, see the textbook of Ref.~\cite{Kovchegov:2012mbw}.}
a particularly striking -- and {\it a priori} surprising -- phenomenon was discovered experimentally at the DESY-HERA collider: {\it hard diffraction}. In a significant proportion of the events (about 10\% overall), the proton left the collision unaltered, while in the forward region of the scattered electron, a hadronic system was observed, as a result of the dissociation of the virtual photon mediating the interaction (see Fig.~\ref{fig-1}).
Diffractive events may be labeled by the size of the region void of particles surrounding the scattered proton, which can be characterized by a Lorentz-invariant {\it rapidity gap} variable~$y_0$. The latter fluctuates from event-to-event
between (almost) zero and the maximum available rapidity $Y=\ln \hat s/Q^2$, where $\hat s$ is the squared center-of-mass energy of the $\gamma^*$-proton/nucleus subreaction, and
$Q$ the virtuality of the photon.
What has been observed in high-energy electron-proton scattering is also expected in electron-nucleus collisions
at a future Electron-Ion Collider (EIC).

 Some time ago, an equation for the distribution of rapidity gaps was rigorously established by Kovchegov and Levin (KL)~\cite{Kovchegov:1999ji} in this context. But solving it analytically remains a formidable challenge.
They did not address directly deep-inelastic scattering, but instead onium-nucleus scattering, which is straightforwardly related to the former when the interaction between the electron and the nucleus is mediated by a {\it longitudinally-polarized} virtual photon.

Let us consider an onium of size $r$ scattering off a big nucleus. In the KL formulation, the distribution of the rapidity gaps
is the solution of a system of two equations. The first one is the Balitsky-Kovchegov (BK) equation for the rapidity evolution of the forward elastic $S$-matrix element.\footnote{The (dimensionless) total, elastic and inelastic cross sections {\it per impact parameter} may be
derived from $S$, which is essentially real at high energy:
$\sigma_{\text{tot}}=2(1-S)$, $\sigma_{\text{el}}=(1-S)^2$, $\sigma_{\text{in}}=\sigma_{\text{tot}}-\sigma_{\text{el}}=1-S^2$.
These formulas show, in particular, that the elastic cross section is maximum (and equal to the inelastic one) when $S=0$.}
Introducing the notation $\bar\alpha\equiv\alpha_s N_c/\pi$, the BK equation reads
\be
\partial_y S(r,y)=\bar\alpha\int\frac{d^2r'}{2\pi}\frac{r^2}{r'^2(r-r')^2}\left[
  S(r',y)S(r-r',y)-S(r,y)
  \right].
\label{eq:S}
\ee
The initial condition is given by e.g. the McLerran-Venugopalan (MV) model,
$S(r,y=0)=e^{-\frac{r^2Q_\text{MV}^2}{4}\ln(e+4/r^2\Lambda_\text{QCD}^2)}$,
with $Q_\text{MV}$ the saturation momentum of the nucleus.
$1/Q_\text{MV}$ can be interpreted as the dipole size
above which the scattering occurs with unit probability [i.e. $S(r\gg1/Q_\text{MV},0)\ll 1$].
The rapidity gap distribution is deduced from an auxiliary function $S_2(r,\tilde y)$
which also obeys the BK equation\footnote{Equations~(\ref{eq:S},\ref{eq:S2})
  also follow quite straightforwardly from the Good-Walker picture,
  see Ref.~\cite{Mueller:2018ned}.}
\be
\partial_{\tilde y} S_2(r,\tilde y)=\bar\alpha\int\frac{d^2r'}{2\pi}\frac{r^2}{r'^2(r-r')^2}\left[
  S_2(r',\tilde y)S_2(r-r',\tilde y)-S_2(r,\tilde y)
  \right],
\label{eq:S2}
\ee
{with the initial condition}
$S_2(r,\tilde y=0)=\left[S(r,y_0)\right]^2$.
In terms of $S_2$, the gap distribution then reads
\be
\frac{d\sigma_{\text{diff}}(y_0|r,Y)}{dy_0}=\left.\frac{\partial}{\partial \tilde y}\right|_{\tilde y=Y-y_0}S_2(r,\tilde y).
\ee

The work presented here may be viewed as an effort to find a solution to the KL set of equations~(\ref{eq:S},\ref{eq:S2}). But instead of trying to solve it brute force, which is technically extremely challenging, we develop a picture of diffractive scattering, from which what we believe should be the asymptotics of the KL equation (almost) straightforwardly follow and which points to a deep link with ancestry problems in branching random walks.
The present write-up shortly summarizes the papers in Refs.~\cite{Mueller:2018ned,Mueller:2018zwx},
before presenting a new numerical comparative study of exact equations for diffraction and ancestry in the dipole model
(see Sec.~\ref{sec:num} below).


\section{Picture of onium-nucleus scattering}

\subsection{Total cross section in the onium and nucleus restframes}

In the onium restframe, the nucleus appears in a highly-evolved and occupied state, while the dominant state of the
onium is a bare quark-antiquark pair.
Event-by-event fluctuations are negligible;
$S$ may be interpreted as the ``transparency'' of the boosted nucleus.

In the nucleus restframe instead, the whole evolution is in the onium, which
appears typically as a set of many gluons (represented by dipoles in the large number-of-color limit~\cite{Mueller:1993rr}),
whose detailed content strongly fluctuates from event-to-event.
For a scattering to occur, there should be at least one gluon in this set
which has a transverse momentum of the order of $Q_\text{MV}$, so
that the whole state has a non-negligible probability to scatter with the nucleus.
In this context, $1-S$ can be interpreted
as the probability that the Fock state of the onium contains at least one gluon with a transverse momentum
of that magnitude.

\subsection{Diffractive cross section in the
  {\bfseries\bf $y_0$}-frame\label{sec:y0frame}}

Let us now choose a frame in which the nucleus is boosted to rapidity~$y_0$
and the onium to rapidity $\tilde y_0=Y-y_0$. In order to have a diffractive event exhibiting a gap~$y_0$
with unit probability, one needs at least
one gluon in the Fock state of the onium
whose transverse momentum is smaller than the saturation scale at rapidity $y_0$, $Q_s(y_0)$.\footnote{%
Denoting by $\chi(\gamma)$ the eigenvalue of the linearized BK equation about $S\sim 1$ corresponding
to the eigenfunction $1-S= r^{2\gamma}$
and by $\gamma_0$ the solution of the equation $\chi(\gamma_0)=\gamma_0\chi'(\gamma_0)$,
one has $Q_s^2(y_0)=Q^2_\text{MV}e^{\bar\alpha y_0\chi'(\gamma_0)}/(\bar\alpha y_0)^{3/2\gamma_0}$.}
Indeed, this condition makes sure that the elastic interaction cross section of the onium is significant.
The diffractive cross section $d\sigma_\text{diff}/dy_0$ is tantamount to this very probability.
A straightforward calculation leads to an elegant formula for the latter, once normalized to the total cross section:
\be
\frac{1}{\sigma_\text{tot}}\frac{d\sigma_\text{diff}}{dy_0}=\text{const}\times
\left[
  \frac{\bar\alpha Y}{\bar\alpha y_0(\bar\alpha Y-\bar\alpha y_0)}
  \right]^{3/2}.
\label{eq:diff_cross_section}
\ee
The overall numerical constant, of order unity, cannot be determined within the present approach.
This formula is actually only valid in the so-called {\it scaling region}, defined by the following
constraints on the parameters:
$1\ll \ln r^2Q_s^2(Y)\ll \sqrt{\chi''(\gamma_0)\bar\alpha Y}$.


\section{Ancestry}

\subsection{Height of genealogical trees in branching random walks}

It has been known for some time that dipole evolution is a peculiar branching random walk~\cite{Munier:2014bba}.
One of the main results of Refs.~\cite{Mueller:2018ned,Mueller:2018zwx} is the surprising observation that the structure of the
branches may be directly
related to an observable in high-energy physics.

Boosting a bare onium of size~$r$ by one unit in the
rapidity~$\tilde y$ opens
the phase space for quantum fluctuations in the form of additional gluons populating its Fock state.
A one-gluon emission by the onium may be interpreted as the splitting of a color dipole into two dipoles, of different
sizes. Upon a further boost, each of these two dipoles may split independently
through the same process.
Thus one understands that QCD evolution is a branching process of dipoles in rapidity,
with a random walk in the sizes of the latter.\footnote{
  The relevant scale for the dipole sizes is logarithmic, and the relevant evolution variable is
  the scaled rapidity $\bar\alpha Y$. Actually, the process is diffusive
  only if one looks at a fixed impact parameter,
  but this is what turns out to be relevant here.}

Now boost to rapidity~$Y$, take e.g. the two largest dipole in the Fock state and
track their first common ancestor. According to Ref.~\cite{0295-5075-115-4-40005},
the rapidity $y_0$ at which the ancestor branches is distributed as
\begin{equation}
p(y_0|r,Y)=
c_p\left[
  \frac{\bar\alpha Y}{\bar\alpha y_0(\bar\alpha Y-\bar\alpha y_0)}
  \right]^{3/2},
\ \text{with}\ c_p=\frac{1}{\bar\gamma}\frac{1}{\sqrt{2\pi\chi''(\gamma_0)}}.
\label{eq:p}
\end{equation}
The value of $\bar\gamma$ depends on which dipoles are picked: In the present case, $\bar\gamma$ coincides with $\gamma_0$.
The formula~(\ref{eq:p}) was actually not established in the peculiar context of dipole evolution of interest
for particle physics, but was argued to
apply to a wide class of branching random walks.
We also expect it to be correct (up to the overall numerical factor) for related quantities, such as
the rapidity distribution of the common
ancestor of {\it all dipoles larger than some given (large enough) size} $1/Q_\text{MV}$.\footnote{%
  This holds true if $r$, $\bar\alpha Y$ and $Q_\text{MV}$ are such that one is in the scaling region
  defined at the end of Sec.~\ref{sec:y0frame}.}

Equation~(\ref{eq:p}) was found by {\it assuming} that the common ancestor was an unusually
large object generated around the rapidity $\tilde y_0=Y-y_0$ in the evolution of the onium~\cite{0295-5075-115-4-40005}.
This is exactly the same mechanism as in the case of the diffraction problem (see Sec.~\ref{sec:y0frame}).
Hence, the two problems are intimately related: up to the overall
normalization, which is determined in the case of the genealogies but
not in the case of diffraction,
$({1}/{\sigma_\text{tot}})({d\sigma_\text{diff}}/{dy_0})$
corresponds to $p(y_0|r,Y)$.

In order to check quantitatively this correspondence between diffraction and ancestry,
we have established exact equations for ancestry
and have compared their numerical solutions to those for diffraction.

\subsection{Ancestry equation for dipoles and its numerical solution\label{sec:num}}

\begin{figure}[t]
  \centering
  \includegraphics[width=.95\textwidth,clip]{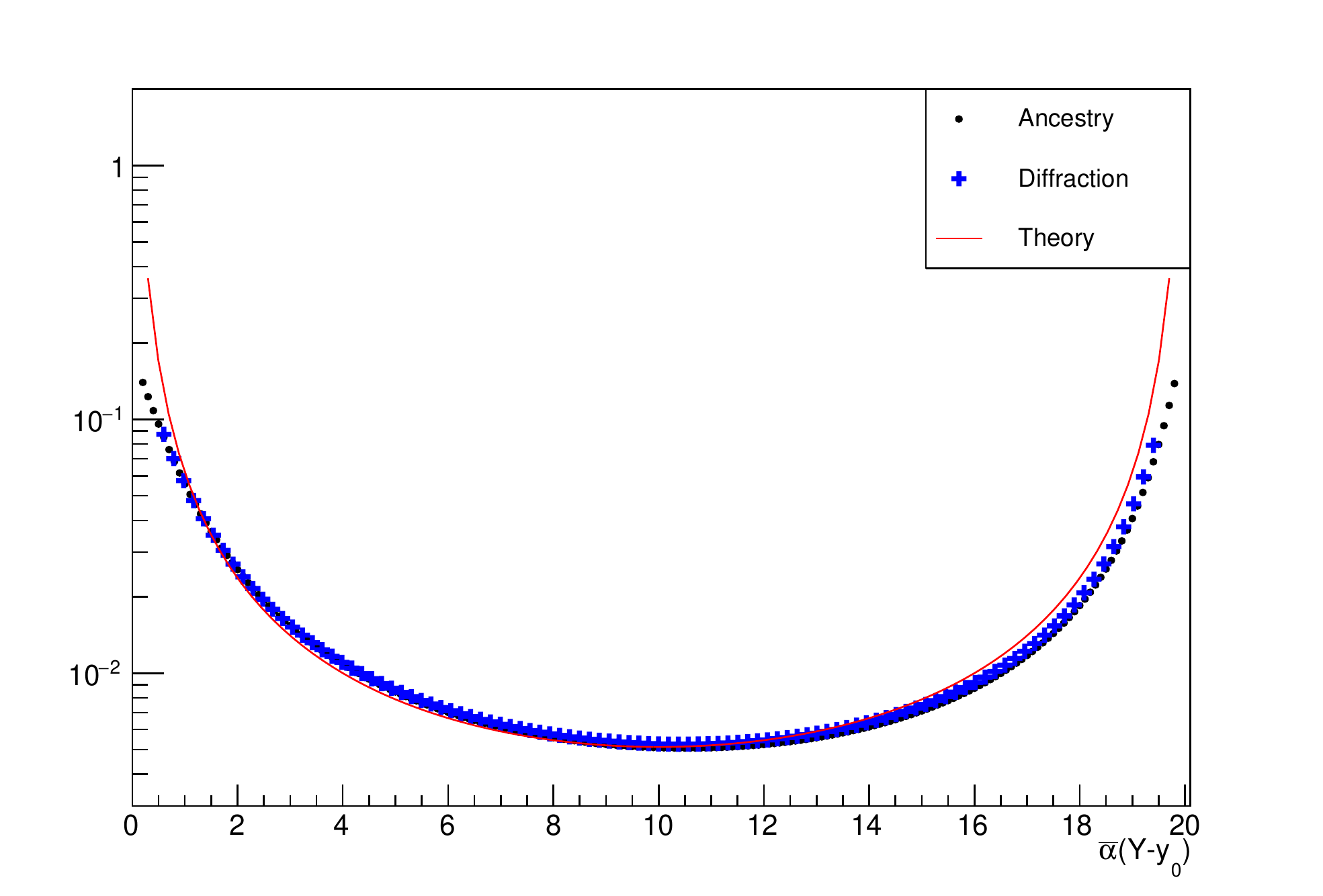}
  \caption{Numerical solution of the equation for
    $({1}/{\sigma_\text{tot}})
    ({d\sigma_\text{diff}}/{dy_0})$
     (labeled ``Diffraction''; from Ref.~\cite{Mueller:2018zwx})
    and for $p_>$ (``Ancestry''; from Ref.~\cite{Dung:2018}) as a function of $\bar\alpha(Y-y_0)$,
    with the parameters set to $\bar\alpha Y=20$ and
    $rQ_\text{MV}=4\times 10^{-21}$.
    The continuous line has been generated from the analytical formula~(\ref{eq:p}) with $\bar\gamma=1$.
  }
\label{fig-2}
\end{figure}

The distribution $p_>(y_0|r,Y)$ of the rapidity at which the first common ancestor
of {all dipoles larger than $1/Q_\text{MV}$} (or, alternatively, of a set of dipoles randomly picked
among the dipoles present in the Fock state at rapidity $Y$ with some probability $T(r)$)
first split obeys the following equation:
\begin{multline}
  \partial_{y} p_>(y_0|r,y)=\bar\alpha\int\frac{d^2r'}{2\pi}\frac{r^2}{r'^2(r-r')^2}\\
  \times\left[
  p_>(y_0|r',y)S(r-r',y)+S(r',y) p_>(y_0|r-r',y)
  -p_>(y_0|r,y)
  \right],
  \label{eq:p>}
\end{multline}
where $S$ solves Eq.~(\ref{eq:S}) with the initial condition $S(r,0)=1-T(r)$.\footnote{If one wanted to consider the common ancestor of all dipoles larger than $1/Q_\text{MV}$, then one would use as an initial condition $T(r,0)=\theta(\ln {r^2Q^2_\text{MV}})$. In our calculation, $T=1-S$ where $S$ is given by the MV model, which is a bit less sharp than the $\theta$ function.}
The initial condition for $p_>$ reads
\be
p_>(y_0|r,y_0)=\bar\alpha\int\frac{d^2 r'}{2\pi}\frac{r^2}{{r'}^2(r-r')^2}[1-S(r',y_0)][1-S(r-r',y_0)].
\label{eq:p>_ic}
\ee
The numerical solutions of the equations for $p_>$ and for $({1}/{\sigma_\text{tot}})
    ({d\sigma_\text{diff}}/{dy_0})$
     are shown in Fig.~\ref{fig-2}:
Both are in very good agreement with Eq.~(\ref{eq:diff_cross_section}), with an overall constant of order~1.
Trying to understand analytically this constant is one of our current goals.



\end{document}